%% file: main.tex
\documentclass[conference]{IEEEtran}
\IEEEoverridecommandlockouts
\usepackage{cite}
\usepackage{amsmath,amssymb,amsfonts}
\usepackage{algorithmic}
\usepackage{graphicx}
\usepackage{textcomp}
\usepackage{xcolor}
\def\BibTeX{{\rm B\kern-.05em{\sc i\kern-.025em b}\kern-.08em
    T\kern-.1667em\lower.7ex\hbox{E}\kern-.125emX}}

\newcommand{\reviseLiu}[1][\textcolor{black}]{#1}

\usepackage{xurl}
\usepackage[colorlinks=true,linkcolor=blue,citecolor=blue,urlcolor=blue]{hyperref}

\usepackage{url}
\usepackage{cleveref}

\usepackage{pifont}

\usepackage{algorithm}
\usepackage{booktabs}
\usepackage{threeparttable}

\usepackage{multirow}

\usepackage{tikz}
\usepackage{listings}
\usetikzlibrary{arrows.meta,positioning,calc}

\definecolor{codebg}{RGB}{246,248,250}
\definecolor{framegray}{RGB}{180,185,190}
\definecolor{titlebg}{RGB}{232,235,239}
\definecolor{darktext}{RGB}{45,50,55}
\lstdefinestyle{yamlbox}{
  basicstyle=\ttfamily\scriptsize,
  backgroundcolor=\color{codebg},
  frame=single,
  rulecolor=\color{framegray},
  xleftmargin=4pt,
  xrightmargin=4pt,
  aboveskip=0pt,
  belowskip=0pt,
  columns=fullflexible,
  keepspaces=true,
  showstringspaces=false,
  breaklines=true,
  breakatwhitespace=false,
  tabsize=2
}

\usepackage{xcolor}
\usepackage{mdframed}

\definecolor{rqgray}{RGB}{242,242,242}

\newmdenv[
  backgroundcolor=rqgray,
  linecolor=black,
  linewidth=2pt,
  topline=false,
  rightline=false,
  bottomline=false,
  leftline=true,
  innerleftmargin=12pt,
  innerrightmargin=12pt,
  innertopmargin=8pt,
  innerbottommargin=8pt,
  skipabove=8pt,
  skipbelow=8pt
]{rqanswer}

\newcommand{\ignore}[1]{}

\usepackage{xspace}
\newcommand{\tool}{\textit{KubeCap}\xspace}
\usepackage{subfigure}

\definecolor{KFTitle}{HTML}{3A3A3A} % 深灰标题栏（接近你参考图）
\definecolor{KFBack}{HTML}{F2F2F2}  % 浅灰正文底
\usepackage[skins,breakable]{tcolorbox}
\usepackage{enumitem} % 控制 itemize 间距和符号
\newtcolorbox{myframebox}[1]{%
  breakable,
  enhanced,
  colback=KFBack,
  boxrule=1.0pt,
  arc=5pt,
  title={#1},
  colbacktitle=KFTitle,
  coltitle=white,
  fonttitle=\bfseries,
  boxed title style={%
    boxrule=0pt,
    arc=5pt,
    left=8pt,right=8pt,
    top=4pt,bottom=4pt
  },
  left=8pt,right=8pt,
  top=8pt,bottom=8pt,
  overlay unbroken={\draw[rounded corners=5pt,line width=1.0pt]
    (frame.south west) rectangle (frame.north east);},
  overlay first={\draw[rounded corners=5pt,line width=1.0pt]
    (frame.south west) rectangle (frame.north east);},
  overlay middle={\draw[line width=1.0pt,color=KFFrame]
    (frame.south west) rectangle (frame.north east);},
  overlay last={\draw[rounded corners=5pt,line width=1.0pt]
    (frame.south west) rectangle (frame.north east);},
    title after break={},
}

\usepackage{microtype}

\begin{document}

\title{\tool: A Framework for Capability Minimization in Kubernetes via Static Analysis and LLM-Assisted Rule Inference}
% {\footnotesize \textsuperscript{*}Note: Sub-titles are not captured for https://ieeexplore.ieee.org  and should not be used}


\author{
\IEEEauthorblockN{
Yuhao Liu,
Yingnan Zhou,
Weijie Liu\textsuperscript{*},
Yan Jia\textsuperscript{*},
and Zheli Liu
}

\IEEEauthorblockA{
Key Laboratory of Data and Intelligent System Security,
Ministry of Education, China (DISSec), \\
Tianjin Key Laboratory of Network and Data Security Technology (NDST), \\
College of Cryptology and Cyber Science, Nankai University, Tianjin 300350, China \\
\{yuhao.liu, yingnan.zhou\}@mail.nankai.edu.cn \\
\{weijieliu, jiay, liuzheli\}@nankai.edu.cn
% \\
% \textsuperscript{$\dagger$}Corresponding authors: Weijie Liu and Yan Jia
}

\thanks{*~Weijie Liu and Yan Jia are corresponding authors.}
}

\maketitle

\input{tex/Section-Abstract}

\input{tex/Section1-Introduction}

\input{tex/Section2-RelatedWork}

\input{tex/Section3-Background}

\input{tex/Section4-EmpiricalStudy}

\input{tex/Section5-Methodology}

\input{tex/Section6-Evaluation}

\input{tex/Section7-Conclusion}

\input{tex/Section-Acknowledgement}

\input{tex/Section-Appendix}

\bibliographystyle{IEEEtran}
\bibliography{bibliography/reference}

% \vspace{12pt}
% \color{red}
% IEEE conference templates contain guidance text for composing and formatting conference papers. Please ensure that all template text is removed from your conference paper prior to submission to the conference. Failure to remove the template text from your paper may result in your paper not being published.

\end{document}

%% file: tex/Section-Abstract.tex
\begin{abstract}
As the most widely used container orchestration platform, Kubernetes provides flexible privilege configuration by allowing developers to manage Linux capabilities via manifest files. However, developers rely on default settings or coarse-grained security contexts in practice, violating the principle of least privilege and enlarging the attack surface of containerized workloads. Existing studies either detect vulnerable patterns in Kubernetes manifests or infer required capabilities for standalone Linux programs, but they do not directly address capability minimization in Kubernetes.

To bridge this gap, we first conduct an empirical study on three open-source datasets, revealing that 74.67\% of projects lack capability configurations. Motivated by our observations, we propose \tool, a framework for Kubernetes capability minimization. \tool translates deployment specifications into deterministic manifests, locates container entrypoints, performs reachability-guided system call analysis, and leverages LLM-assisted rule specification to derive syscall--parameter--capability relations from Linux kernel code. Based on these results, \tool infers the minimal capability set required by each workload and automatically generates repaired manifests. Evaluation on 10 representative Go-based Kubernetes projects shows an average capability reduction rate of 54.97\%, outperforming rapid type analysis and class hierarchy analysis baselines while maintaining practical analysis cost. These results demonstrate \tool's effectiveness in enforcing least privilege in Kubernetes.

%Our results reveal that explicit capability customization remains uncommon. Specifically, 74.67\% of container-related projects lack capability configurations entirely, and only 19.33\% adopt a least-privilege pattern.

\end{abstract}


%% file: tex/Section1-Introduction.tex
%-------------------------------------------------------------------------------
\section{Introduction}
%-------------------------------------------------------------------------------
%\footnote{Graduated and Incubating Projects, \url{https://www.cncf.io/projects/}}

Kubernetes (K8s) is an open source container orchestration system governed by Cloud Native Computing Foundation~(CNCF)~\cite{cncf-projects}. It is used to automate the deployment, capacity expansion, load balancing, rolling update, fault recovery, and other tasks of container applications~\cite{2016K8s}. Kubernetes applications are typically deployed through declarative YAML manifests, which specify workload objects such as Pods, which are the basic execution units that encapsulate one or more containers. Among the configuration fields of the manifests, \texttt{securityContext} defines security-sensitive runtime settings for a container or Pod. The \texttt{securityContext} field in a Pod manifest controls permission-related parameters, such as \texttt{capabilities}, \texttt{privileged}, and \texttt{runAsNonRoot}, which directly determine how much authority a containerized process holds over the underlying host system.

\begin{figure}
\begin{center}
\includegraphics[width=0.8\hsize]{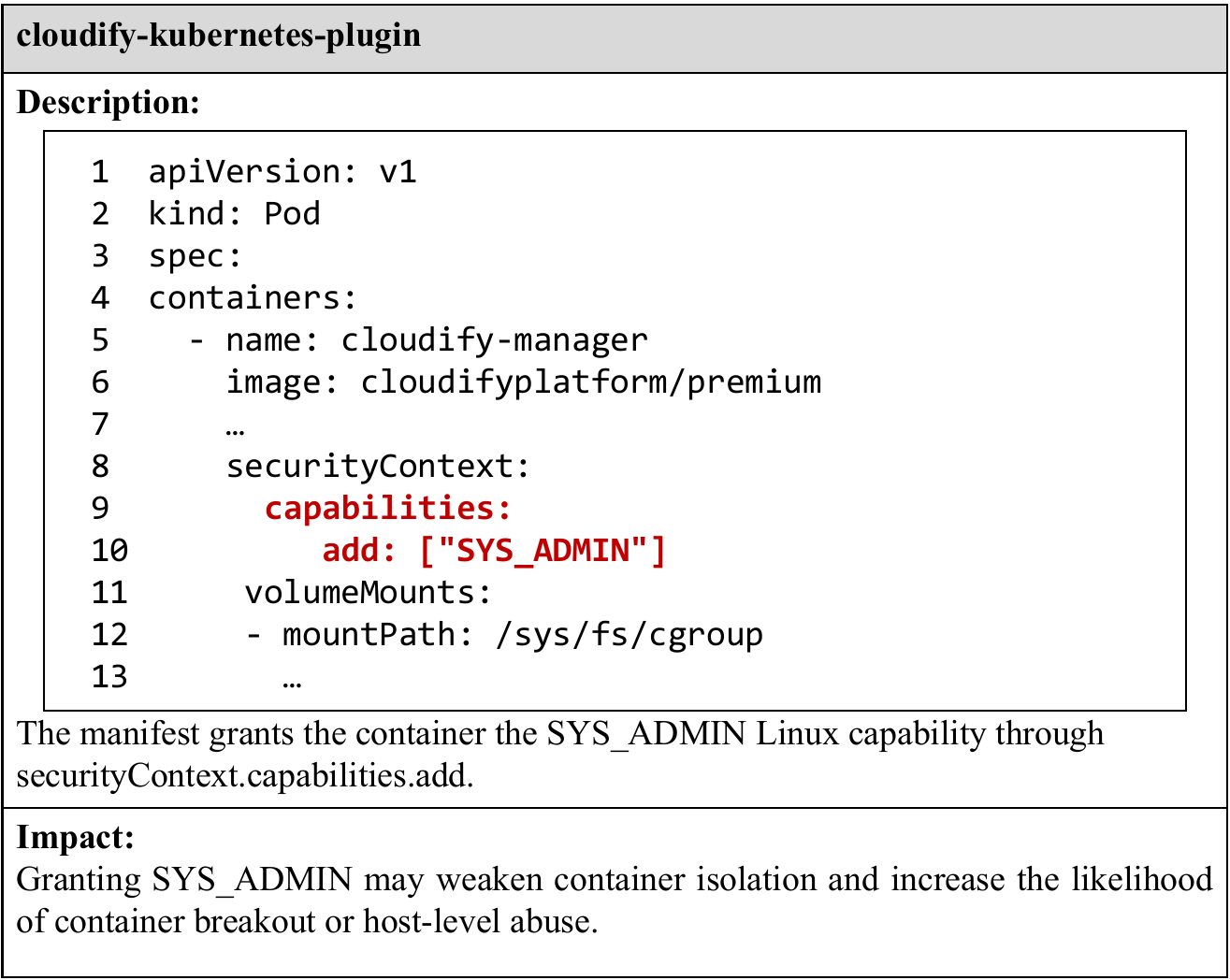}
\end{center}
\caption{\label{fig_1} An example of the privileged container in the pod manifest of cloudify-kubernetes-plugins.}
\end{figure}

Although Kubernetes has become the mainstream platform for container orchestration, its flexibility in configuring container privileges introduces severe security challenges~\cite{2023eBPF,2018USENIXNamespace}. The mechanisms of \texttt{securityContext} are intended to support the principle of least privilege. Security guidelines and compliance controls recommend minimizing assigned capabilities whenever possible~\cite{cis-c220}. However, the least privilege is difficult to enforce in practice. In Kubernetes deployments, to avoid complex permission debugging, developers often grant containers excessive privileges in the \texttt{securityContext} of YAML manifests, such as directly assigning \texttt{CAP\_SYS\_ADMIN} or enabling \texttt{privileged: true}~\cite{2023tosem-empiricalstudy}. Even when developers do not explicitly enable fully privileged mode, containers still inherit a default set of Linux capabilities assigned by the container runtime, many of which are unnecessary for the application’s normal operation. This over-privileged configuration greatly expands the system's attack surface. For example, as shown in Figure~\ref{fig_1}, the manifest grants the container \texttt{cloudify-manager} the broad \texttt{CAP\_SYS\_ADMIN} capability. Such a configuration weakens container isolation and exposes the workload to a larger set of kernel-level operations. This excessive privilege may further increase the risk of container escape or host-level abuse. In addition, He et al.~\cite{2023eBPF} and Yang et al.~\cite{2023takeover} demonstrated that even a single misconfigured privileged container can compromise an entire cluster through eBPF-based privilege escalation.

To bridge the gap, we conducted an empirical study on three open-source Kubernetes datasets. Our results reveal that explicit capability configuration remains uncommon. Specifically, 74.67\% of projects lack capability configurations, and only 19.33\% adopt the least-privilege pattern. Based on these findings, we propose \tool, a Kubernetes capability minimization framework with five components. First, it translates raw K8s template files into deployable manifests and extracts workload-level capability configurations. Second, it performs entrypoint localization to recover the actual executables and source-level entry files of containers, including multi-container settings such as init containers. Third, it conducts reachability-guided system call analysis on the recovered entrypoints to identify the system calls that are relevant to the deployed workload. Fourth, it leverages the LLM-assisted rule specification and minimal capability inference to derive precise syscall--parameter--capability relations from Linux kernel code and infer the minimal capability set required by each workload. Finally, it performs deviation analysis between the originally declared capabilities and the inferred minimal set, and automatically generates repaired manifests that enforce least privilege.

% We implement \tool as a practical prototype on top of \texttt{ar-go-tools} and use \texttt{GPT-4o-mini} for rule extraction. 

We evaluate \tool on 10 representative Go-based projects of Kubernetes. The results show that \tool achieves an average capability reduction rate of 54.97\%, removing redundant privileges from real-world workloads. Compared with rapid type analysis (RTA) and class hierarchy analysis (CHA), the reachability-based design achieves a higher average reduction rate (54.97\%, 38.00\% and 7.68\%, respectively). Points-to analysis~(PTA) is competitive when it completes but often times out on larger projects. Case studies further validate the practicality of \tool on real-world projects. In addition, \tool achieves practical performance that the analysis completes within a few minutes on the largest subjects and within tens of seconds on most others, while keeping memory usage in a similar range of about 1 to 3.5 GiB. These results demonstrate that \tool can effectively and efficiently precompute minimized capabilities for K8s workloads.

In summary, we make the main contributions as follows:
\begin{itemize}

    \item \textbf{Empirical study.} 
    We conducted an empirical study on three open-source Kubernetes datasets and show that 74.67\% of projects do not explicitly configure capabilities, \reviseLiu{and only 19.33\% explicitly remove default capabilities.}
     % We find that 74.67\% of projects do not explicitly configure capabilities.

     \item \textbf{Tool.} We propose \tool, a framework for Kubernetes capability minimization. It can infer workload-specific minimal capability sets and generate repaired K8s manifests to prevent the over-privileged capability configurations in Kubernetes manifests. We make \tool publicly available at
    {\url{https://github.com/anabioticsoul/KubeCap/}}.

    \item \textbf{Effectiveness.} Evaluation on 10 Kubernetes projects shows that \tool removes redundant capabilities, achieving an average reduction rate of 54.97\% with practical time and memory overhead.

\end{itemize}

% Recently, Rahman et al.~\cite{2023tosem-empiricalstudy} point out the harmfulness of absent or privileged \texttt{securitycontext}. They also built a tool SLI-KUBE to detect misconfigurations. 
% Meanwhile, Cesarano et al.~\cite{2025kubefence} and Sgan Cohen et al.~\cite{2025kubeguard} can suggest generic “safe” configurations but lack behavioral validation and thus risk breaking legitimate workloads.
% System-level analyses like Decap~\cite{2022Decap} derive capability requirements for standalone binaries, yet fail to capture container orchestration semantics.

%% file: tex/Section2-RelatedWork.tex
\section{Related Work}
\reviseLiu{Prior studies have explored least-privilege enforcement from different perspectives. 
One line of work focuses on detecting security misconfigurations in Kubernetes manifests. 
Rahman et al.~\cite{2023tosem-empiricalstudy} systematically mined open-source Kubernetes projects and showed that security misconfigurations are widespread in real-world deployments. 
They further proposed SLI-KUBE, a rule-based analysis tool that uses Def-Use chain analysis to trace key configuration parameters across multiple Kubernetes manifest files. 
Malul et al.~\cite{2024genkubesec} collected and labeled a large corpus of Kubernetes configuration files, established a unified misconfiguration index, and fine-tuned an LLM to detect potential misconfigurations. 
These studies are effective at identifying vulnerable configuration patterns in Kubernetes manifests. 
However, they focus on whether a configuration violates known best practices, rather than whether the granted privileges are functionally required by the deployed workload. }

\reviseLiu{Another line of work infers required capabilities for privilege reduction. A key challenge here is that capability checks are often deeply embedded in helper functions and triggered only under specific syscall parameters. While Confine~\cite{2020Confine} restricts attack surfaces via \texttt{seccomp} filtering, tools like Decap~\cite{2022Decap} and LiCA~\cite{2022Lica} directly infer capabilities for privileged programs through static and path-sensitive analysis. Runtime solutions like CAPMAN~\cite{2025Capman} mitigate capability abuses, but their reliance on behavior learning, kernel instrumentation and stable workloads limits their applicability in dynamic Kubernetes environments. Furthermore, due to the complexity of whole-kernel static analysis~\cite{SPLC20BDD}, prior approaches typically rely on manual rule extraction covering limited capabilities (e.g., \texttt{CAP\_SYS\_ADMIN})~\cite{2022Decap}. Crucially, they mainly target standalone binaries or generic images, failing to account for Kubernetes-specific complexities such as rendered manifests, command overrides, and init containers. To address these limitations, \tool performs demand-driven, LLM-assisted rule extraction. By tracing bounded caller chains from capability-checking sites, \tool leverages an LLM to summarize syscall-level parameter conditions into structured rules, enabling a highly accurate mapping of both direct and condition-dependent capability requirements.}

\reviseLiu{Recent K8s works focus on API-layer and manifest hardening. 
KubeFence~\cite{2025kubefence} reduces the Kubernetes API attack surface by deriving fine-grained API filtering policies from K8s workloads. 
KubeGuard~\cite{2025kubeguard} uses LLMs with manifests and runtime logs to recommend least-privilege K8s configurations, including Roles, NetworkPolicies, and Deployments. 
Different from these systems, \tool focuses specifically on Linux capability minimization in container \texttt{securityContext} fields. 
It links rendered manifests, actual container entrypoints, reachable syscall sites, and kernel-derived syscall--parameter--capability rules to precompute workload-specific capability sets. 
Therefore, \tool is complementary to existing Kubernetes misconfiguration detectors, capability-inference tools, and hardening systems.}

%% file: tex/Section3-Background.tex
\section{Background}

% \subsection{Kubernetes Architecture}
% Kubernetes is the de facto platform for orchestrating containerized applications, providing automated support for deployment, scaling, scheduling, and recovery across distributed environments~\cite{2016K8s}. Its control plane coordinates workload placement and lifecycle management, while the underlying container runtime executes application containers on cluster nodes. This abstraction greatly simplifies cloud-native application management, but also introduces security risks. In particular, Kubernetes workloads may require heterogeneous privileges, access shared kernel resources, and run in multi-tenant environments, making fine-grained privilege control critical for reducing attack surfaces and preventing unauthorized host-level operations~\cite{2018USENIXNamespace,2023eBPF}.

\subsection{Capabilities in Kubernetes}
Linux capabilities provide a fine-grained privilege model by decomposing superuser privileges into independently enabled or disabled capability units~\cite{linux_capabilities_man7}. In Kubernetes, developers customize container privileges through the \texttt{securityContext} field using \texttt{capabilities.add} and \texttt{capabilities.drop}~\cite{k8s_security_context}. Kubernetes itself does not define a fixed default capability set for containers; instead, the effective privilege set depends on the underlying container runtime~\cite{k8s_container_runtimes,k8s_security_context}. Therefore, if a container does not explicitly drop unnecessary capabilities, it may retain a non-trivial baseline capability set granted at runtime.

% The Linux Capabilities mechanism was originally designed to break down traditional root privileges into fine-grained permission units.
% Linux capabilities split ``root power'' into finer privileges, and are widely used in containers as a compromise between ``all-powerful root'' and fully unprivileged processes. [7] In container environments, Docker documents a default set of allowed capabilities (including NET\_RAW), and explains that --privileged grants all capabilities and broadly removes confinement (devices, AppArmor/SELinux). [8] In practice, this means a Kubernetes Pod that does not explicitly drop capabilities often inherits a non-trivial baseline privilege set from the underlying runtime. [9]

\begin{table}[h]
\centering
% \footnotesize

\caption{Representative risks of excessive capabilities in Kubernetes}
\label{tab:cap-threats}
\resizebox{0.9\columnwidth}{!}{
\begin{tabular}{@{}lll@{}}
\toprule
\textbf{Capability} & \textbf{CVEs} & \textbf{Impact} \\ 
\midrule
CAP\_NET\_RAW
& CVE-2020-14386~\cite{CVE-2020-14386}
& Packet forgery \\

CAP\_NET\_ADMIN
& CVE-2017-6074~\cite{CVE-2017-6074}
& Container compromise \\

CAP\_SYS\_ADMIN
& CVE-2022-0185~\cite{CVE-2022-0185}
&  Cluster compromise \\
% CAP\_SYS\_ADMIN & CVE-2016-4557 & container escalation \\ 
\bottomrule
\end{tabular}
}
\end{table}

\begin{table*}[th]
\centering
% \footnotesize
\caption{Project-level prevalence of capability-related configurations across three OSS Kubernetes datasets.}
\label{tab:rq1_dataset_compare}
\resizebox{0.75\hsize}{!}{
\begin{tabular}{@{}lccccc@{}}
\toprule
\textbf{Dataset} & \textbf{Total} & \textbf{w/ Container} & \textbf{w/ SecurityContext} & \textbf{w/ Capabilities} & \textbf{w/ Drop-All Capabilities} \\
\midrule
Rahman et al.~\cite{2023tosem-empiricalstudy}    & 57 & 52 & 19 (36.5\%) & 10 (17.3\%) & 7 (13.5\%) \\
Shamim et al.~\cite{2025icse-empiricalstudy}     & 33 & 15 & 10 (46.7\%) & 4 (26.7\%)  & 2 (13.3\%) \\
This work & 102 & 83 & 51 (61.4\%) & 24 (28.9\%) & 20 (24.1\%) \\
\midrule
\textbf{Total} & 192 & 150 & \textbf{53.33\%} & \textbf{25.33\%} & \textbf{19.33\%} \\
\bottomrule
\end{tabular}
}
\end{table*}

\subsection{Threat Model}
\begin{figure}[h]
\centering
\includegraphics[width=\linewidth]{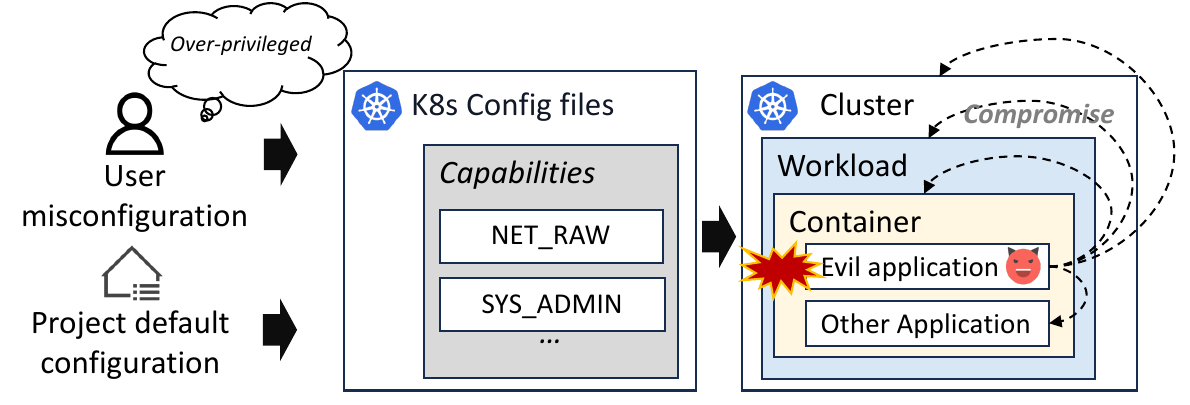}
\caption{How excessive capability configurations in Kubernetes amplify post-compromise risks.}
\label{fig:overprivileged-example}
\end{figure}

% TODO展示案例与过度权限的表格
In Kubernetes, the security risk of Linux capabilities does not arise merely from their existence, but from the gap between the capabilities actually required by a workload and those retained by default or granted in configuration. Whether inherited from the container runtime or explicitly added in YAML manifests, unnecessary capabilities enlarge the attack surface available to an attacker after initial container compromise. As shown in \Cref{fig:overprivileged-example}, there are two aspects that can make over-privileged capabilities especially risky in Kubernetes.

\textbf{Container escalation.} 
Once a container is compromised, its privileges may lead to host-level impact. The K8s documentation notes that, without Linux user namespaces, a root container may obtain root privileges on the node after a breakout, and any capabilities granted to the container remain valid on the host as well~\cite{k8sdoc-userspace}. Note that K8s namespaces and Linux namespaces are different. K8s namespaces are API-level abstractions for organizing Kubernetes objects, while Linux namespaces isolate user and group identities between a container and the host. Moreover, as shown in Table~\ref{tab:cap-threats}, certain kernel attack surfaces are directly controlled or facilitated by specific capabilities. For example, the OSS-security advisory for CVE-2020-14386 states that creating AF\_PACKET sockets requires \texttt{CAP\_NET\_RAW} in the corresponding network namespace. Therefore, retaining capabilities such as \texttt{CAP\_NET\_RAW} can expose containers to a broader class of kernel networking attacks, such as packet forgery. More generally, prior measurement studies have shown that default container configurations are often overly permissive. Lin et al.~\cite{2018measurement} found that a substantial fraction of real-world exploits could still be successfully launched from within containers under default settings, and further showed that kernel security mechanisms such as capabilities, seccomp, and MAC play a more critical role in preventing privilege escalation than namespaces and cgroups alone~\cite{CVE-2020-14386,CVE-2017-6074}.

\textbf{Compromising the whole cluster.} In Kubernetes, excessive capabilities may impact an entire cluster beyond a single container or node. Once an attacker gains code execution in an over-privileged container, those capabilities can be used to achieve container escape or node-level compromise. Table~\ref{tab:cap-threats} shows that broad capabilities attacks, such as CVE-2022-0185, are associated with cluster-level compromise risks. Recent work has shown that after escaping a container via eBPF-enabled privileges (typically requiring \texttt{CAP\_SYS\_ADMIN}), attackers can abuse over-privileged Operator Pods and their ServiceAccounts on the same node to interact with the K8s API server and create or update Pods on other nodes, thereby expanding the attack from one compromised node to the entire cluster~\cite{2023eBPF}. Note that Operator Pods are Kubernetes controllers. Unlike standard application Pods, which execute business logic, Operator Pods are designed to watch, create, update, or reconcile Kubernetes resources via the API server. This threat model assumes that the attacker has already obtained initial execution in a container and attempts to exploit capabilities that exceed the workload’s actual needs to compromise the node and pivot through over-privileged Kubernetes components.

% To address these challenges, Kubernetes provides the \texttt{SecurityContext} mechanism, which governs the security-relevant properties of Pods and containers, including user and group identities, Linux capabilities, privilege escalation permissions, and file system isolation semantics. SecurityContext operates at two hierarchical levels. \ding{172} PodSecurityContext applies pod-wide settings such as supplemental groups or file system ownership. \ding{173} SecurityContext, which configures security behaviors for an individual container, allowing precise tailoring of each component’s privilege model. The overarching goal of these mechanisms is to ensure that containers execute with the least privileges necessary, thereby reducing the risk of container breakout, unauthorized host access, data exposure, and privilege misuse. When properly configured, SecurityContext serves as a critical defense layer that restricts the operational authority of workloads, enforces isolation boundaries, and strengthens the overall security posture of Kubernetes deployments.

%% file: tex/Section4-EmpiricalStudy.tex
\section{Empirical study: Prevalence of Over-privileged Capabilities in Real-world Kubernetes Projects}\label{sec3}

We first conduct an empirical study of how Linux capabilities are configured in real-world projects. We collect and examine capability-related configurations from open-source Kubernetes repositories. Our study aims to answer two questions. 1) How frequently are capabilities explicitly configured in real-world K8s projects? 2) What level of granularity are the capabilities configured in practice? In particular, do developers apply least-privilege patterns such as dropping all the capabilities and adding back only those necessary, or do they retain broad defaults, add capabilities in an ad hoc manner, or even use privileged containers? If explicit configuration is uncommon, or if projects mainly rely on default or overly coarse-grained privilege settings, this would indicate substantial room for improvement. Answering these questions helps us understand the current practice of capability configuration and motivates the need for privilege analysis and recommendation.
% Before designing a tool to minimize excessive privileges in Kubernetes security contexts, w

\subsection{Study Design and Dataset Construction}

% To understand whether Linux capabilities are explicitly configured in practice, we conduct an empirical study on open-source Kubernetes projects. 

\textbf{Search source.}
To broaden the empirical basis of our study, we incorporate two public datasets released by prior work on Kubernetes configurations. The first is the dataset of Rahman et al.~\cite{2023tosem-empiricalstudy}, which investigates security misconfigurations in open-source Kubernetes manifests. The second is the dataset of Shamim et al.~\cite{2025icse-empiricalstudy}, which analyzes community-prescribed security configurations for Kubernetes. Since these datasets were originally constructed for different research goals, we do not directly reuse their reported statistics. Instead, we re-analyze all included projects under a unified standard. To complement these datasets, we construct our own dataset from GitHub, which provides publicly accessible deployment artifacts and has been widely used in previous studies.
% open-source repositories provide

\textbf{Search strategy.}
Following prior studies that mined GitHub repositories for software engineering research~\cite{2016emseMining,2017emseGithub}, we used ``kubernetes helm'' as the search query and systematically crawled candidate repositories. We applied explicit filtering rules in line with common empirical dataset construction practices~\cite{2017fseOSP}. Specifically, we excluded repositories with fewer than 20 stars, filtered out archived repositories, and removed inaccessible or duplicate repositories. This process resulted in a dataset of 102 open-source repositories.

% This normalization enables a consistent cross-dataset comparison of capability-related configuration practices in RQ1.

\begin{figure*}[th]
\begin{center}
\includegraphics[width=0.8\hsize]{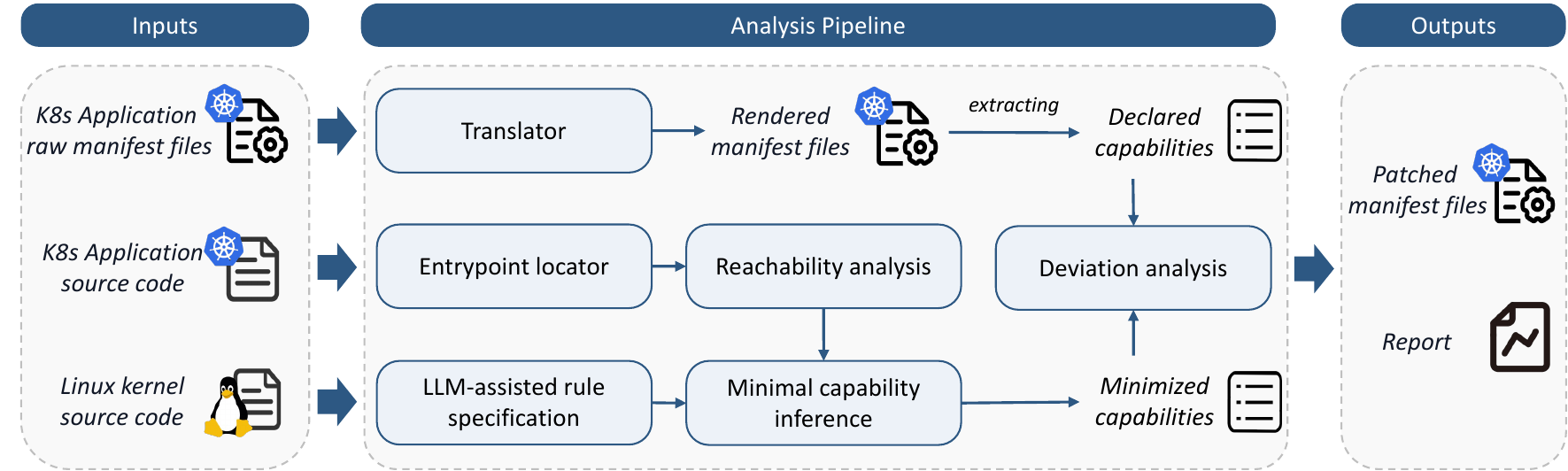}
\end{center}
\caption{The workflow of \tool.}\label{fig:workflow}
\end{figure*}

\textbf{Quantification analysis.}
We analyzed and identified the Kubernetes manifests in projects that configure containers (hereafter referred to as \textit{container projects}). For each project, we examined whether at least one manifest explicitly configures \texttt{securityContext}, \texttt{capabilities}, or the least-privilege pattern \texttt{drop: ["ALL"]} (i.e., drop all capabilities). We used a single project rather than an individual YAML file as the unit of analysis, so each repository was counted at most once per category. To reduce extraction errors, we manually validated a subset of parsed manifests and confirmed the consistency between the extracted fields and the original YAML files.

% \textbf{Analysis Result.}
% Our analysis shows that explicit capability customization is still uncommon in real-world Kubernetes projects. Among the 102 repositories we collected, 83 contain container-related Kubernetes manifests. However, only 51 of these projects explicitly configure \texttt{securityContext}, and only 24 further specify \texttt{capabilities}. Even fewer projects, 20 in total, adopt the stronger least-privilege pattern \texttt{drop: ["ALL"]}. These results suggest that, although container security settings are not entirely ignored, fine-grained capability minimization is still not common practice. In many projects, developers appear to rely on default runtime settings or coarse-grained privilege controls rather than explicitly restricting capabilities for each workload.

% \subsection{RQ1: Prevalence of Over-privileged Capabilities in Real-world Kubernetes Projects}

\subsection{Prevalence of Explicit Capabilities Configuration}
\reviseLiu{Table~\ref{tab:rq1_dataset_compare} summarizes the prevalence of capability-related configurations across the three datasets.}
%For each project, we identify whether it contains container-related Kubernetes manifests, and whether at least one manifest explicitly configures \texttt{securityContext}, \texttt{capabilities}, or \texttt{drop: ["ALL"]}. Each project is counted at most once per category.

\textbf{Observation 1:}~\textit{Capability configuration is uncommon across all three datasets. 112 of 150 projects (74.67\%) do not set capabilities and thus still rely on Kubernetes default configuration, while only 19.33\% drop all capabilities.}
\Cref{tab:rq1_dataset_compare} summarizes the results. In Rahman et al.'s dataset, only 10 of 52 container projects (17.3\%) configure \texttt{capabilities}, and only 7 (13.5\%) use the drop-all pattern. In Shamim et al.'s dataset, 4 of 15 container projects (26.7\%) configure \texttt{capabilities}, and only 2 (13.3\%) adopt \texttt{drop: ["ALL"]}. Our dataset shows the same trend that among 83 container projects, only 24 (28.9\%) configure \texttt{capabilities}, and only 20 (24.1\%) use the drop-all pattern. These results indicate that explicit capability customization remains uncommon in real-world Kubernetes projects.

\textbf{Observation 2:}~\textit{The configuration of security context is more common than capabilities. Across the three datasets, 53.33\% of container projects specify security context, but only 25.33\% explicitly configure capabilities.}
As shown in \Cref{tab:rq1_dataset_compare}, this gap is consistent across all three datasets. In Rahman et al.'s dataset, 19 of 52 container projects (36.5\%) define \texttt{securityContext}, but only 10 (17.3\%) further configure \texttt{capabilities}. In Shamim et al.'s dataset, 10 of 15 projects (46.7\%) specify \texttt{securityContext}, compared with 4 (26.7\%) that configure \texttt{capabilities}. Our dataset shows that 51 of 83 projects (61.4\%) specify \texttt{securityContext}, while only 24 (28.9\%) explicitly configure \texttt{capabilities}. These results suggest that developers are more likely to adopt coarse-grained security settings than fine-grained ones.

Overall, the results suggest that least-privilege capability configuration is still not common practice in real-world Kubernetes projects. This gap motivates the need for capability analysis and minimization.

%% file: tex/Section5-Methodology.tex
\section{Methodology}
\subsection{Overview}
We propose \tool which is designed for minimizing over-privileged capability configurations in Kubernetes. As shown in Figure~\ref{fig:workflow}, it first translates raw template files into Kubernetes manifests, locates the container runtime entrypoints, and performs reachability-guided system call analysis on the corresponding source code. In parallel, \tool extracts syscall--condition--capability rules from Linux kernel code via the LLM-assisted rule specification. It then combines both results to infer the minimal capability set required by the workload, compares it with the original manifest, and generates a repaired least-privilege configuration.

% 3. \tool Methodology
% To achieve automated privilege reduction, \tool designs a complete technical pipeline including configuration parsing, Intermediate Representation (IR) conversion, reachability analysis, and deviation evaluation.

\subsection{Configuration Translator}
In modern cloud-native practice, Kubernetes workloads are rarely specified as plain static YAML files. Instead, developers often define workloads using Helm charts and Kustomize overlays~\cite{helmdoc,k8sdoc-kustomize} to dynamically generate manifests from templates, dependency packages, and environment-specific values. Directly analyzing these raw templates would lose the deployment semantics and lead to inaccurate privilege inference. To address this problem, \tool introduces the \emph{Configuration Translator} that translates raw template files into deterministic Kubernetes manifests. Given a target repository, the translator identifies its configuration paradigm in the order of Helm charts, Kustomize overlays, and plain YAML files. For Helm-based repositories, \tool recursively discovers candidate chart directories by locating \texttt{Chart.yaml} files, while excluding common testing fixtures such as \texttt{testdata} and \texttt{tests}. For each chart, \tool invokes a resilient rendering procedure that first executes \texttt{helm dependency build} and then runs \texttt{helm template} to generate the fully expanded manifest. The rendering output is persisted as a fixed intermediate artifact, \texttt{rendered.yaml}, which serves as the canonical input to subsequent analysis stages. To improve robustness, \tool performs bounded automatic repair and retry when rendering fails. Specifically, it handles several common failure modes observed in practice, including invalid placeholder versions in \texttt{Chart.yaml}, malformed \texttt{apiVersion} fields, missing chart dependencies, and absent \texttt{.Values.*} parameters. Temporary metadata changes are reverted after rendering so that the original repository state is preserved. 
%This design substantially improves the renderability of noisy open-source charts while keeping the analysis reproducible and non-intrusive.

\begin{figure}[t]
\centering
%%% 图片大小
\begin{tikzpicture}[font=\small, scale=1, every node/.style={transform shape}]

% ---------- top ----------
\node[anchor=north west, inner sep=0pt] (topcode) at (0,0) {
\begin{minipage}{0.88\linewidth}
\begin{lstlisting}[style=yamlbox]
containers:
- name: {{ .Chart.Name }}
  image: "{{ .Values.image.repository }}:{{ .Values.image.tag }}"
  imagePullPolicy: {{ .Values.image.pullPolicy }}
  command:
    - ./podinfo
    - --port={{ .Values.service.httpPort | default 9898 }}
    - --cert-path={{ .Values.tls.certPath }}
    - --port-metrics={{ .Values.service.metricsPort }}
    - --grpc-port={{ .Values.service.grpcPort }}
    - --grpc-service-name={{ .Values.service.grpcService }}
    - --level={{ .Values.logLevel }}
...
\end{lstlisting}
\end{minipage}
};

\node[
  anchor=south west,
  fill=titlebg,
  text=darktext,
  font=\bfseries\footnotesize,
  rounded corners=2pt,
  inner xsep=6pt,
  inner ysep=3pt
] at ([xshift=2pt,yshift=2pt]topcode.north west) {Helm Template};

% ---------- bottom ----------
\node[anchor=north west, inner sep=0pt] (botcode) at ([yshift=-1.2cm]topcode.south west) {
\begin{minipage}{0.88\linewidth}
\begin{lstlisting}[style=yamlbox]
containers:
- name: podinfo
  image: "ghcr.io/stefanprodan/podinfo:6.9.4"
  imagePullPolicy: IfNotPresent
  command:
    - ./podinfo
    - --port=9898
    - --cert-path=/data/cert
    - --port-metrics=9797
    - --grpc-port=9999
    - --grpc-service-name=podinfo
    - --level=info
...
\end{lstlisting}
\end{minipage}
};

\node[
  anchor=south west,
  fill=titlebg,
  text=darktext,
  font=\bfseries\footnotesize,
  rounded corners=2pt,
  inner xsep=6pt,
  inner ysep=3pt
] at ([xshift=2pt,yshift=2pt]botcode.north west) {Rendered Manifest};

% ---------- arrow ----------
\draw[-{Latex[length=3.5mm,width=2.5mm]}, line width=0.9pt, darktext]
($(topcode.south)+(0,-0.25cm)$) -- ($(botcode.north)+(0,0.25cm)$);

\node[
  fill=white,
  text=darktext,
  font=\footnotesize\itshape,
  inner xsep=4pt,
  inner ysep=1pt
] at ($(topcode.south)!0.5!(botcode.north)$)
{Helm rendering};

\end{tikzpicture}
\caption{An example of Helm template rendering.}
\label{fig:helm-rendering-example}
\end{figure}
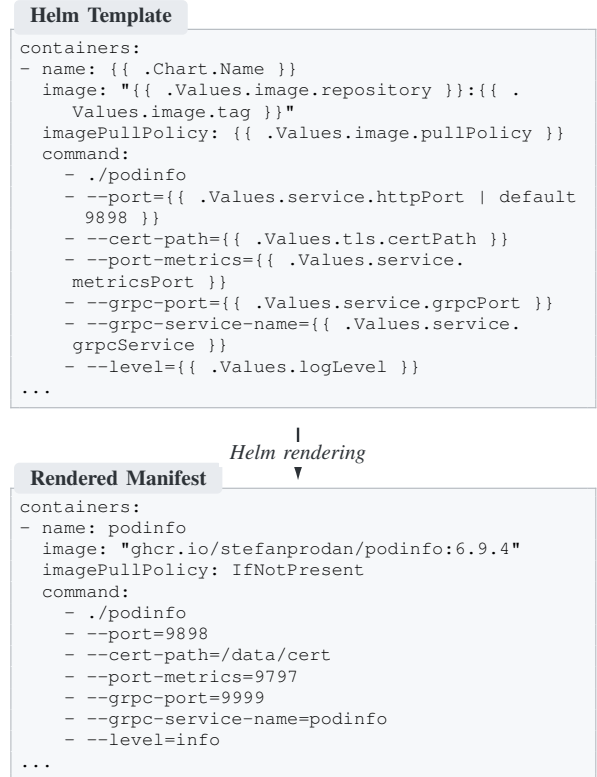

After rendering, the translator parses \texttt{rendered.yaml} to identify workload objects, including \textit{Deployment}, \textit{DaemonSet}, \textit{StatefulSet}, \textit{Job}, \textit{CronJob}, and \textit{Pod}. For each workload, it extracts the \texttt{securityContext} together with container definitions, distinguishing \texttt{containers} from \texttt{initContainers}. It also records the declared image, \texttt{command}, and \texttt{args} fields, which are essential to recover the actual runtime entrypoint in the next stage. For example, as shown in \Cref{fig:helm-rendering-example}, \tool renders the \textit{podinfo} Helm template into a deterministic \texttt{rendered.yaml} fragment by resolving variables in the container specification. In the original template, fields such as container name, image name, image pull policy, and command-line arguments are parameterized with Helm expressions (e.g., \texttt{\{\{ .Chart.Name \}\}} and \texttt{\{\{ .Values.* \}\}}). After rendering, these symbolic placeholders are materialized into concrete runtime values, such as \texttt{podinfo}, \texttt{ghcr.io/stefanprodan/podinfo:6.9.4}, \texttt{IfNotPresent}, and explicit command arguments like \texttt{--port=9898} and \texttt{--grpc-port=9999}. 

\subsection{Entrypoint Locator}
While Kubernetes manifests specify container images, static analysis requires the exact program entrypoint and the corresponding source code file. In practice, this mapping is obscured by multiple layers of indirection, including image defaults, manifest overrides, and wrapper scripts.

To correlate rendered manifests with the source code of K8s application, we design the entrypoint locator. It takes the manifests as input, parsing each workload from \texttt{rendered.yaml} and iterating containers. For each container, \tool pulls the specified image, retrieves the default entrypoints and arguments (e.g., \texttt{command} and \texttt{args}) recorded in the metadata. It then combines the image defaults with the arguments according to K8s command resolution semantics to recover the runtime startup command. However, the startup command may not correspond to the actual application entrypoint. Many containers use wrapper scripts or helper binaries (e.g., \texttt{sh -c}) for environment initialization and signal handling. \tool therefore applies heuristic script unwrapping to recursively resolve such indirections until it reaches the underlying executable responsible for the main application logic. Once the target executable is identified, \tool extracts the binary from the container and maps it back to the source repository. 

Since many Kubernetes components and cloud-native controllers are implemented in Go, \tool leverages Go-specific binary analysis to bridge compiled artifacts and source code. In particular, it uses \texttt{go tool} to analyze symbol information and locate the \texttt{main} function. By matching the recovered file paths with the cloned repository structure, \tool pinpoints the exact entry file associated with the rendered container configuration. The output of this stage is a precise mapping from each rendered Kubernetes workload to its actual executable entrypoint and source code path. It ensures that static single assignment (SSA) construction and reachability analysis are performed on the exact code paths induced by the rendered Kubernetes workload, rather than on unrelated binaries, dead code, or repository-wide over-approximations.
% This mapping is critical for the subsequent analysis pipeline. 

\subsection{System Call Analysis based on Graph Reachability}
After obtaining the source code path at container runtime through the entrypoint locator, the system needs to calculate all system calls that the program may trigger during actual execution. Since modern cloud-native projects have massive standard libraries and third-party dependencies, directly performing a full scan at the source code level would lead to an overwhelming number of false positives.

\reviseLiu{To address this problem, \tool performs reachability-guided system call analysis on top of the SSA representation.} As shown in Algorithm~\ref{alg:reach_syscall}, \tool first initializes the entry functions from the target \texttt{main} package and then iteratively expands the worklist to compute the set of reachable functions. For each visited function, \tool first performs syscall classification and then discovers its callees from SSA instructions. Newly discovered callees are added to the worklist until a fixed point is reached. In this way, \tool restricts the analysis scope to functions that are actually reachable from the real container entrypoint. \tool reachability analyzer traverses the reachable graph top-down and identifies three types of system call trigger points:
% , rather than indiscriminately scanning the entire codebase

\textbf{Syscall traps.} Direct invocations of low-level trap instructions, such as \texttt{syscall.Syscall} and \texttt{syscall.Syscall6}.

\textbf{Syscall wrappers.} Secure encapsulations of the operating system within the high-level language's standard library, such as various file and network operations under the \texttt{golang.org/x/sys/unix} package.

\textbf{Syscall runtimes.} Low-level calls triggered by the language runtime for garbage collection and goroutine scheduling, such as \texttt{futex}- and \texttt{mmap}-related calls in the \texttt{runtime} package.

By strictly restricting the analysis to reachable paths originating from the true entrypoint, this module naturally filters out unlinked and unexecuted dead code within the project. Consequently, it outputs a precise set of system calls tightly bound to the specific K8s configuration context.

\begin{algorithm}[t]
\small
\caption{Reachability-Guided System Call Analysis}
\label{alg:reach_syscall}
\begin{algorithmic}[1]

\REQUIRE SSA program $P$, entry options $\mathit{excludeMain}$ and $\mathit{excludeInit}$
\ENSURE Reachable functions $R$ and syscall categories $C_{trap}, C_{wrap}, C_{rt}$

\STATE $E \gets$ initialize entry functions in package \texttt{main}
\STATE $R \gets E$; $W \gets E$
\STATE $C_{trap}, C_{wrap}, C_{rt} \gets \emptyset$

\WHILE{$W \neq \emptyset$}
    \STATE $f \gets$ pop($W$)
    \IF{\textsc{IsTrap}$(f)$}
        \STATE $C_{trap} \gets C_{trap} \cup \{\textit{fullName}(f)\}$
    \ELSIF{\textsc{IsRuntimeSyscall}$(f)$}
        \STATE $C_{rt} \gets C_{rt} \cup \{\textit{fullName}(f)\}$
    \ELSIF{\textsc{IsWrapper}$(f)$}
        \STATE $C_{wrap} \gets C_{wrap} \cup \{\textit{fullName}(f)\}$
    \ENDIF

    \STATE $S \gets$ \textsc{FindCallees}$(f)$
    \FOR{each $g \in S$}
        \IF{$g \notin R$}
            \STATE $R \gets R \cup \{g\}$
            \STATE push($W, g$)
        \ENDIF
    \ENDFOR
\ENDWHILE

\STATE \textbf{return} $(R, C_{trap}, C_{wrap}, C_{rt})$

\end{algorithmic}
\end{algorithm}

\subsection{LLM-assisted Rule Specification}
\label{sec:llm_rule_specification}
\reviseLiu{\textbf{Kernel slice extraction.}
\tool first constructs capability-relevant kernel code slices. It identifies representative capability-checking APIs in the Linux kernel, such as \texttt{capable()}, \texttt{ns\_capable()}, and related variants. Each matched location is treated as a seed capability-check site. For each seed site, \tool performs bounded caller tracing to recover the syscall-level context of the capability check. Starting from the function that contains the capability-checking API, \tool recursively searches its callers until a syscall-like function is reached or the caller context no longer contributes syscall-level parameter information. The resulting slice contains not only the capability-checking call itself, but also the relevant caller context, branch predicates, and argument-propagation information needed to explain when the check can be triggered.}

\reviseLiu{\textbf{Rule Inference.}
\tool employs one-shot learning~\cite{2006Oneshot}, a technique widely adopted in prior studies~\cite{2020Fewshot}. The prompt consists of four components: task description, input specification, output format, and a representative example. First, the task instruction defines the LLM as a Linux kernel and security expert, and provides a detailed description of the task, including the types of inputs provided, how they should be processed, and the required structure and content of the output. Second, \tool provides three types of inputs: (1) \textit{Source file}, (2) \textit{Line range}, and (3) \textit{Caller-augmented code slice}. The output is required to be a structured array containing: (1) \textit{Syscall name}; (2) \textit{Parameter information}; (3) \textit{Trigger condition}; (4) \textit{Inferred capability}; and (5) \textit{Side condition}. If the provided slice does not support a valid syscall--parameter--capability relationship, the LLM must return an empty array. Third, the representative example enables the LLM to generate responses following the required output schema. The generated rules are subsequently consumed by \tool's deterministic minimal capability inference stage, which matches them against reachable syscall sites and resolved argument constraints in the analyzed workload.}

\subsection{Minimal Capability Inference}
The rules extracted from kernel code are not directly sufficient for capability minimization, because many capability requirements are guarded by argument-dependent predicates rather than triggered unconditionally. Therefore, after constructing the syscall-condition-capability rule base, \tool performs a dedicated constraint resolution stage to determine which conditional capability rules are actually satisfiable along the reachable execution paths of the target workload.

Given the set of reachable functions derived from the true container entrypoint, \tool processes all conditional rules associated with the observed system calls. For each rule, it first identifies the argument positions involved in the condition. These argument indices may either be explicitly encoded in the rule metadata or extracted from the condition expression itself. \tool then attempts to resolve the corresponding argument values through a constant analysis procedure. Concretely, it prioritizes direct numeric literals at call sites, then queries previously resolved constants from the analysis cache, followed by source-level constant definition lookup, and finally performs SSA-based backward tracing to recover constant origins. The resolved values are cached in the analysis manager to avoid redundant propagation across rules and call sites.

To improve precision for frequently used syscall wrappers, \tool further incorporates a set of syscall-specific handlers. For representative interfaces such as \texttt{fcntl} and \texttt{ioctl}, these handlers directly inspect wrapper call sites, retrieve concrete arguments from callers, and match them against rule predicates. Once a condition is satisfied, the corresponding conditional capability requirement is appended to the inferred result. This design allows \tool to handle common argument-sensitive system calls more precisely than purely generic propagation. At the same time, for highly context-sensitive interfaces such as \texttt{open}, \texttt{mmap}, and \texttt{ptrace}, \tool deliberately avoids overly aggressive approximation in the special path in order to reduce false positives.

After all rule conditions have been resolved, \tool aggregates the matched capability requirements across the reachable execution scope and computes the minimized capability set, denoted as $Cap_{min}$. Intuitively, $Cap_{min}$ contains only those Linux capabilities whose guarding conditions are actually exercised by the program under the analyzed Kubernetes configuration. In this way, \tool bridges the gap between syntactic syscall reachability and semantically justified privilege requirements, enabling capability minimization to be driven by both program behavior and argument-level constraints.

\subsection{Deviation Analysis}
After inferring the minimized capability set required by the program, \tool compares the analysis result with the declaration in Kubernetes configuration. Specifically, \tool compares the original capability set declared in the manifest, denoted as $Cap_{orig}$, with the minimized capability set inferred from program analysis, denoted as $Cap_{min}$. It then computes the redundant capability set as:
\[
Cap_{del} = Cap_{orig} \setminus Cap_{min}.
\]
Each capability in $Cap_{del}$ represents a privilege granted in the original \texttt{securityContext} but not required by the actual workload behavior, and is therefore treated as over-privileged. Based on this deviation analysis, \tool automatically generates a patched Kubernetes manifest to enforce least privilege. \tool explicitly inserts \texttt{drop: ["ALL"]} into the \texttt{securityContext} to eliminate inherited default privileges. In addition, it adds back only the minimized capability set $Cap_{min}$ through the \texttt{add: [...]} field.

%% file: tex/Section6-Evaluation.tex
\section{Evaluation}
We evaluate \tool by answering the following research questions:

% from three perspectives. First, we assess its effectiveness in identifying and minimizing redundant capabilities on 10 representative Go-based Kubernetes-related projects, including both overall reduction results and a real-world case study. Second, we conduct an ablation study to compare different analysis strategies and examine how they affect capability reduction effectiveness. Third, we evaluate the efficiency and scalability of \tool by measuring the time and memory overhead of these analysis strategies on the same set of projects. Through these experiments, we aim

\begin{table*}[t]
\centering
\caption{Effectiveness of \tool in Reducing Capabilities.}
\label{tab:effectiveness}
\resizebox{0.8\hsize}{!}{%
\begin{tabular}{@{}l l c c c c@{}}
\toprule
\textbf{Project} & \textbf{Container} & \textbf{Original Caps (${Cap}_{orig}$)} & \textbf{LoC} & \textbf{Reduced Caps (${Cap}_{del}$)} & \textbf{Reduction (\%)} \\
\midrule
amazon-vpc-cni-k8s & aws-node & 2 & 8,861,258 & 1  & 50.00 \\
deckhouse & caps-controller-manager & 14 & 2,206,710 & 6  & 42.86 \\
eks-anywhere & kube-vip & 2 & 17,179,855 & 1  & 50.00 \\
podinfo & podinfo & 14 & 924,637 & 10 & 71.43 \\
twitter-go & feeds & 14 & 1,489,482 & 7  & 50.00 \\
ced & ced-server & 14 & 1,029,028 & 10 & 71.43 \\
gardener-extension-cri-resmgr & gardener-extension-cri-resmgr & 14 & 5,784,166 & 6  & 42.86 \\
Intel-ECI-ExCat-K8s & csl-excat-deviceplugin & 41 & 1,666,339 & 35 & 85.37 \\
minibroker & minibroker & 14 & 3,553,638 & 6  & 42.86 \\
poolprovider-for-k8s & k8s-poolprovider & 14 & 1,294,652 & 6  & 42.86 \\
\midrule
\textbf{Average} & -- & -- & -- & -- & \textbf{54.97} \\
\bottomrule
\end{tabular}
}
\end{table*}

% \textbf{RQ1:} How prevalent are over-privileged capability configurations in real-world Kubernetes projects?
% , and what risks do they introduce

\textbf{RQ1:} How effective is \tool in identifying and minimizing over-privileged capabilities in real-world Kubernetes workloads?

\textbf{RQ2:} How do different call-graph construction strategies affect \tool’s capability reduction effectiveness?

\textbf{RQ3:} What is the runtime performance of \tool for large-scale Kubernetes analysis?

\subsection{Experiment Setup}
\noindent\textbf{Implementation.} We implement \tool on top of ar-go-tools~\cite{ar-go-tools}, a static analysis framework for Go that offers reusable infrastructure for program loading, SSA-based analysis, data-flow reasoning, and reachability analysis. \tool leverages GPT-4o-mini~\cite{gpt-4o-mini} to infer the syscall-condition-capability specifications. All experiments were conducted on a server with two Intel Xeon Gold 6248R CPUs at 3.00~GHz, 64~GB RAM, and a 1~TB hard disk, running Ubuntu 18.04 LTS and Python 3.6.8. 

\noindent\textbf{Dataset.} For the evaluation, we selected 10 representative Go-based Kubernetes-related projects from the three datasets introduced in \Cref{sec3}, covering both CNCF projects and Helm-based OSS projects: \textit{amazon-vpc-cni-k8s}~\cite{amazon-vpc-cni-k8s}, \textit{deckhouse}~\cite{deckhouse}, \textit{eks-anywhere}~\cite{eks-anywhere}, \textit{podinfo}~\cite{podinfo}, \textit{twitter-go}~\cite{twitter-go}, \textit{ced}~\cite{ced}, \textit{gardener-extension-cri-resmgr}~\cite{gardener-extension-cri-resmgr}, \textit{Intel-ECI-ExCat-K8s}~\cite{intel-eci-excat-k8s}, \textit{minibroker}~\cite{minibroker}, and \textit{poolprovider-for-k8s}~\cite{poolprovider-for-k8s}. We focus on Go projects because the current implementation of \tool is built on the Go toolchain. Restricting the subjects to Go therefore ensures that all compared methods operate under the same language setting and analysis pipeline.

 % (48 physical cores / 96 hardware threads in total)
 
% \textbf{Existing dataset setup.}

% \textbf{GitHub dataset setup.}
% In addition to the two public datasets, we construct our own dataset from GitHub to characterize real-world capability configuration practices in Helm-oriented Kubernetes projects. Using \texttt{kubernetes helm} as the primary search query, we systematically crawled candidate repositories and applied several filtering rules to improve dataset quality. Specifically, we excluded repositories with fewer than 20 stars to remove personal test projects and low-quality samples, filtered out archived repositories to retain actively maintained projects, and removed inaccessible or duplicate repositories. After this process, we obtained 102 open-source repositories. We then analyzed their YAML files and retained the projects that contain container-related Kubernetes manifests as the valid subjects for capability-configuration analysis. For each remaining project, we applied the same project-level statistics as above. 
% This dataset serves as the primary empirical basis of our RQ1 study., i.e., whether the project contains container workloads and whether at least one manifest explicitly specifies \texttt{securityContext}, \texttt{capabilities}, or \texttt{drop: ["ALL"]}

% For each project, we compare our reachability-based pipeline with three widely used call-graph construction strategies, namely CHA, RTA, and VTA, in terms of total analysis time and average peak heap memory usage.

\subsection{RQ1: Effectiveness of \tool}
We examine whether \tool can identify and remove redundant capabilities from real-world Kubernetes workloads. \reviseLiu{In addition, we examine syscall-site and kernel capability analysis to understand how many rules can be resolved with high confidence and how many require conservative approximation. Appendix~\ref{app:A} provides the detailed analysis.}

\subsubsection{Overall reduction effectiveness}
We evaluate the effectiveness of \tool by comparing the capability set originally declared in Kubernetes manifests ($Cap_{orig}$) with the minimized capability set inferred by our analysis ($Cap_{min}$). For each workload, \tool identifies capabilities that are declared but not required by the reachable execution paths and removes them from the final policy. Table~\ref{tab:effectiveness} summarizes the effectiveness of \tool in removing redundant capabilities for 10 representative Go-based projects of Kubernetes.

Overall, \tool consistently identifies excessive privileges across all evaluated subjects, achieving an average reduction rate of 54.97\%. For projects that originally use the default Kubernetes capability set, \tool removes between 6 and 10 capabilities, corresponding to reduction rates from 42.86\% to 71.43\%, indicating that a substantial fraction of the originally granted privileges are unnecessary. For example, \tool removes 6 capabilities (42.86\%) from \textit{deckhouse}, 7 capabilities (50\%) from \textit{twitter-go}, and 10 capabilities (71.43\%) from both \textit{podinfo} and \textit{ced}. Even for projects with smaller original capability sets, \tool is still able to eliminate redundant privileges: it removes 1 out of 2 capabilities for both \textit{amazon-vpc-cni-k8s} and \textit{eks-anywhere}, corresponding to a 50\% reduction. Notably, for \textit{Intel-ECI-ExCat-K8s}, whose container is originally configured with \texttt{privileged: true}, \tool identifies 35 removable capabilities, corresponding to an 85.37\% reduction. This result shows that even highly over-privileged workloads can be substantially tightened. Overall, the results indicate that real-world Kubernetes workloads often request more privileges than necessary, and that \tool can effectively identify and remove such redundant capabilities.

\begin{table*}[t]
\centering
\caption{Comparison of Capability Reduction Results Across Reachability, RTA, CHA, and PTA.}
\label{tab:capability-comparison}
\resizebox{0.9\hsize}{!}{%
\begin{tabular}{@{}l ccc ccc ccc ccc@{}}
\toprule
\multirow{2}{*}{\textbf{Project}} &
\multicolumn{3}{c}{\textbf{Reachability}} &
\multicolumn{3}{c}{\textbf{RTA}} &
\multicolumn{3}{c}{\textbf{CHA}} &
\multicolumn{3}{c}{\textbf{PTA}} \\
\cmidrule(lr){2-4} \cmidrule(lr){5-7} \cmidrule(lr){8-10} \cmidrule(lr){11-13}
& \textbf{\#Nodes} & \textbf{${Cap}_{del}$} & \textbf{Reduction (\%)}
& \textbf{\#Nodes} & \textbf{${Cap}_{del}$} & \textbf{Reduction (\%)}
& \textbf{\#Nodes} & \textbf{${Cap}_{del}$} & \textbf{Reduction (\%)}
& \textbf{\#Nodes} & \textbf{${Cap}_{del}$} & \textbf{Reduction (\%)} \\
\midrule
amazon-vpc-cni-k8s & 129,616 & 1  & 50.00 & 143,939 & 0  & 0.00  & 189,919 & 0  & 0.00  & N/A    & N/A & N/A \\
deckhouse & 33,831 & 6  & 42.86 & 41,530 & 6  & 42.86 & 47,238 & 1  & 7.14  & N/A    & N/A & N/A \\
eks-anywhere & 59,947 & 1  & 50.00 & 47,620 & 0  & 0.00  & 61,606 & 0  & 0.00  & N/A    & N/A & N/A \\
podinfo & 14,301 & 10 & 71.43 & 15,614 & 6  & 42.86 & 18,282 & 1  & 7.14  & 10,913 & 6   & 42.86 \\
twitter-go & 11,324 & 7  & 50.00 & 11,296 & 7  & 50.00 & 13,699 & 1  & 7.14  & 7,846  & 6   & 42.86 \\
ced & 8,990 & 10 & 71.43 & 8,464  & 8  & 57.14 & 14,432 & 1  & 7.14  & 4,621  & 6   & 42.86 \\
gardener-extension-cri-resmgr & 51,871 & 6  & 42.86 & 57,421 & 6  & 42.86 & 64,263 & 1  & 7.14  & N/A    & N/A & N/A \\
Intel-ECI-ExCat-K8s & 30,896 & 35 & 85.37 & 36,811 & 24 & 58.54 & 40,799 & 11 & 26.83 & N/A    & N/A & N/A \\
minibroker & 37,396 & 6  & 42.86 & 41,761 & 6  & 42.86 & 43,310 & 1  & 7.14  & N/A    & N/A & N/A \\
poolprovider-for-k8s & 22,583 & 6  & 42.86 & 27,372 & 6  & 42.86 & 31,067 & 1  & 7.14  & N/A    & N/A & N/A \\
\midrule
\textbf{Average} & -- & \textbf{8.8} & \textbf{54.97} & -- & \textbf{6.9} & \textbf{38.00} & -- & \textbf{1.8} & \textbf{7.68} & -- & N/A & N/A \\
\bottomrule
\end{tabular}
}
\begin{tablenotes}
\scriptsize
\item[] *N/A indicates that no valid result was obtained.
\end{tablenotes}
\end{table*}

\subsubsection{Case studies}
% Functional Verification of Minimized Manifests

\reviseLiu{To evaluate whether the capabilities minimized by \tool remain practical for real-world projects, we validate the patched manifests on representative Kubernetes applications and check whether their intended functionality is preserved after removing the capabilities identified as unnecessary. We conducted case studies on three representative projects from three datasets respectively, covering different Kubernetes usage scenarios (cloud-native applications and Helm-based projects), including \textit{amazon-vpc-cni-k8s}, \textit{podinfo}, and \textit{minibroker}. The \textit{amazon-vpc-cni-k8s} project is one of the most widely used Kubernetes CNI plugins in AWS environments and is suitable for validation because its manifest contains security-sensitive privilege settings while its correctness directly affects inter-Pod networking. The \textit{podinfo} project exposes diverse cloud-native interfaces, including HTTP, WebSocket, gRPC, metrics, storage, cache, and service-to-service communication, making it suitable for testing user-visible application behavior across different deployment styles. The \textit{minibroker} project implements the Open Service Broker workflow on Kubernetes and exercises Helm-based provisioning, Kubernetes resource labeling, secret/service discovery, credential generation, and deprovisioning. For each subject, we deployed both the original manifest and the patched manifest, exercised a documentation-driven functional workload, and compared whether the minimized version preserved the same externally observable behavior as the original version.}
% The \textit{amazon-vpc-cni-k8s} project is one of the most widely used Kubernetes CNI plugins in AWS environments. This project is particularly suitable for validation because its manifest contains highly security-sensitive privilege settings, including Linux capabilities and privileged execution, while its correctness directly affects inter-Pod networking. 

\reviseLiu{For project \textit{amazon-vpc-cni-k8s}, the original manifest explicitly requested security-sensitive privileges, including \texttt{CAP\_NET\_ADMIN}, \texttt{CAP\_NET\_RAW}, and privileged execution. \tool identified \texttt{CAP\_NET\_RAW} as removable for \texttt{aws-node}, while \texttt{CAP\_NET\_ADMIN} and privileged execution were retained because removing them caused readiness failures or \textit{``CrashLoopBackOff''}. We validated the patched manifest by checking CNI health and Pod-to-Pod connectivity. We provide the detailed case study in Appendix~\ref{app:B}. For \textit{podinfo}, the original manifests did not explicitly configure capabilities and therefore inherited the runtime default capability set. We validated three deployment scenarios: a standalone deployment, a three-tier stack, and a webapp-style two-service deployment. The workload covered representative operations such as rollout, liveness probes, HTTP APIs, service-to-service communication, etc. For \textit{Minibroker}, \tool used patched manifest and preserved the existing non-root and read-only-root-filesystem settings. We validated the complete Open Service Broker lifecycle across five built-in service classes, including \texttt{mariadb}, \texttt{mongodb}, \texttt{mysql}, \texttt{postgresql}, and \texttt{redis}. The workload exercised key stages such as catalog generation, provisioning, Helm installation, secret discovery, etc. After minimization, \textit{amazon-vpc-cni-k8s} remained healthy and preserved inter-Pod connectivity, \textit{podinfo} passed all tested deployment and application-level checks, and \textit{Minibroker} completed the full broker lifecycle for all five service classes. These results suggest that the removed capabilities were unnecessary for the functionality, while capabilities required for correct execution were conservatively retained. The detailed verification results can be found in our repository.}

\begin{rqanswer}
\textbf{Answer to RQ1:} \tool effectively removes redundant capabilities from real-world Kubernetes workloads. On 10 representative Go-based projects, \tool achieves an average reduction rate of 54.97\%. It eliminates 42.86\% to 71.43\% of default capabilities across most subjects.
\end{rqanswer}

\subsection{RQ2: Ablation study}
\reviseLiu{We next study how different analysis strategies affect the reduction results of \tool. We include class hierarchy analysis (CHA)~\cite{1995CHA}, rapid type analysis (RTA)~\cite{1996RTA}, and points-to analysis (PTA)~\cite{1996PTA} as representative call-graph construction baselines widely used in static analysis. The reachability analysis used in \tool is an SSA-based analysis that resolves both direct calls and interface dispatches using SSA-level type information associated with call sites. Compared with traditional reachability algorithms~\cite{1998reachability}, it achieves higher precision in call resolution. Unlike CHA and RTA, it further leverages SSA-level interface construction to identify potential dynamic dispatch targets. Compared with PTA, it does not perform full points-to tracking or interprocedural data-flow analysis, thereby trading marginal precision for essential scalability while remaining practically sound. \Cref{tab:capability-comparison} compares the capability reduction results produced by these four analysis strategies on 10 projects. For each project, we report the number of analyzed nodes, the number of removable capabilities ($Cap_{del}$), and the corresponding reduction rate.}
 % (which relies solely on class hierarchy relations)  (which restricts call targets to instantiated types)

Overall, Reachability achieves the strongest capability reduction among the four strategies, with an average of 8.8 removable capabilities (a 54.97\% reduction rate), compared to 6.9 (38.00\%) for RTA and only 1.8 (7.68\%) for CHA. Reachability outperforms CHA across all projects and exceeds RTA on several key subjects, including \textit{amazon-vpc-cni-k8s}, \textit{eks-anywhere}, \textit{podinfo}, \textit{ced}, and \textit{Intel-ECI-ExCat-K8s}. Crucially, the results demonstrate that analyzing more nodes does not necessarily lead to better capability reduction. Although CHA generally covers the largest number of nodes, it removes the fewest capabilities. This suggests that CHA's coarse over-approximation retains substantially more unnecessary privilege requirements. In contrast, Reachability often analyzes fewer nodes while achieving stronger reductions, indicating that its entrypoint-guided pruning can better focus the analysis on code that is actually relevant to the deployed container.

\reviseLiu{To further understand the impact of this over-approximation, we conducted a case study on \textit{ced}, comparing the capability sets derived from Reachability and RTA under the same workload. The reachability-based analysis resulted in 8 capabilities, whereas RTA resulted in 15 capabilities. Functional testing confirmed that both configurations preserved all documented functionality in \textit{ced}, including CSV data transfer, group management operations, server liveness checks, fuzzy search, and RSVP updates. This indicates that RTA’s coarser approximation retains redundant capabilities, while Reachability achieves a more precise, functionally sound least-privilege configuration.}

While PTA shows competitive reduction results on the few subjects where it completes, its applicability is severely limited in our evaluation. As shown in \Cref{tab:capability-comparison}, PTA frequently exceeded the timeout threshold on large real-world entrypoints and is therefore reported as N/A. Based on our investigation, these N/A results stem from a complexity spike in pointer-based call-graph construction rather than an analysis failure. The PTA branch triggers whole-program Andersen-style pointer analysis with reflection, and its computational cost increases sharply when the entrypoint induces a much larger SSA and reachable-function scope. The long silent periods started after computing the call graph, indicating that the dominant overhead lies inside call-graph construction. Overall, these results confirm that Reachability provides the optimal trade-off between analysis scope, scalability, and capability minimization effectiveness.

\begin{rqanswer}
\textbf{Answer to RQ2:} Reachability achieves the strongest capability reduction overall, with an average reduction rate of 54.97\%. It outperforms RTA (38.00\%) and CHA (7.68\%). While PTA can be competitive on a few completed cases, it often fails to finish on larger real-world entrypoints.
\end{rqanswer}

% Overall, this case study demonstrates that \tool can provide practically meaningful privilege minimization results.

% \begin{table}[t]
% \centering
% \caption{Dataset Characteristics}
% \label{tab:dataset}
% \begin{tabular}{cccc}
% \toprule
% \textbf{Project} & \textbf{Stars}  & \textbf{Workloads} & \textbf{LOCs}\\
% \midrule
% Nginx-ingress         & 15.2K                       & 4                   & 250K            \\

% \bottomrule
% \end{tabular}
% \begin{tablenotes}
% \centering
% \scriptsize
% \item[] Project (Abbr. for Project name); \\
% Stars (Abbr. for Github stars); \\
% Workloads (Abbr. for Kubernetes workloads); \\ 
% LOCs  (Abbr. for Lines of Codes).
% \end{tablenotes}
% \end{table}

% \subsection{RQ2: Effectiveness of \tool}
% \subsubsection{Real-world}

\subsection{RQ3: Efficiency and Scalability of \tool}
\reviseLiu{To evaluate the practicality of \tool, as shown in ~\Cref{fig:efficiency}, we measure its total analysis time and average peak heap memory across the 10 representative projects. }

\subsubsection{Time overhead}
\begin{figure}[h]
  \centering
    \includegraphics[width=1\hsize]{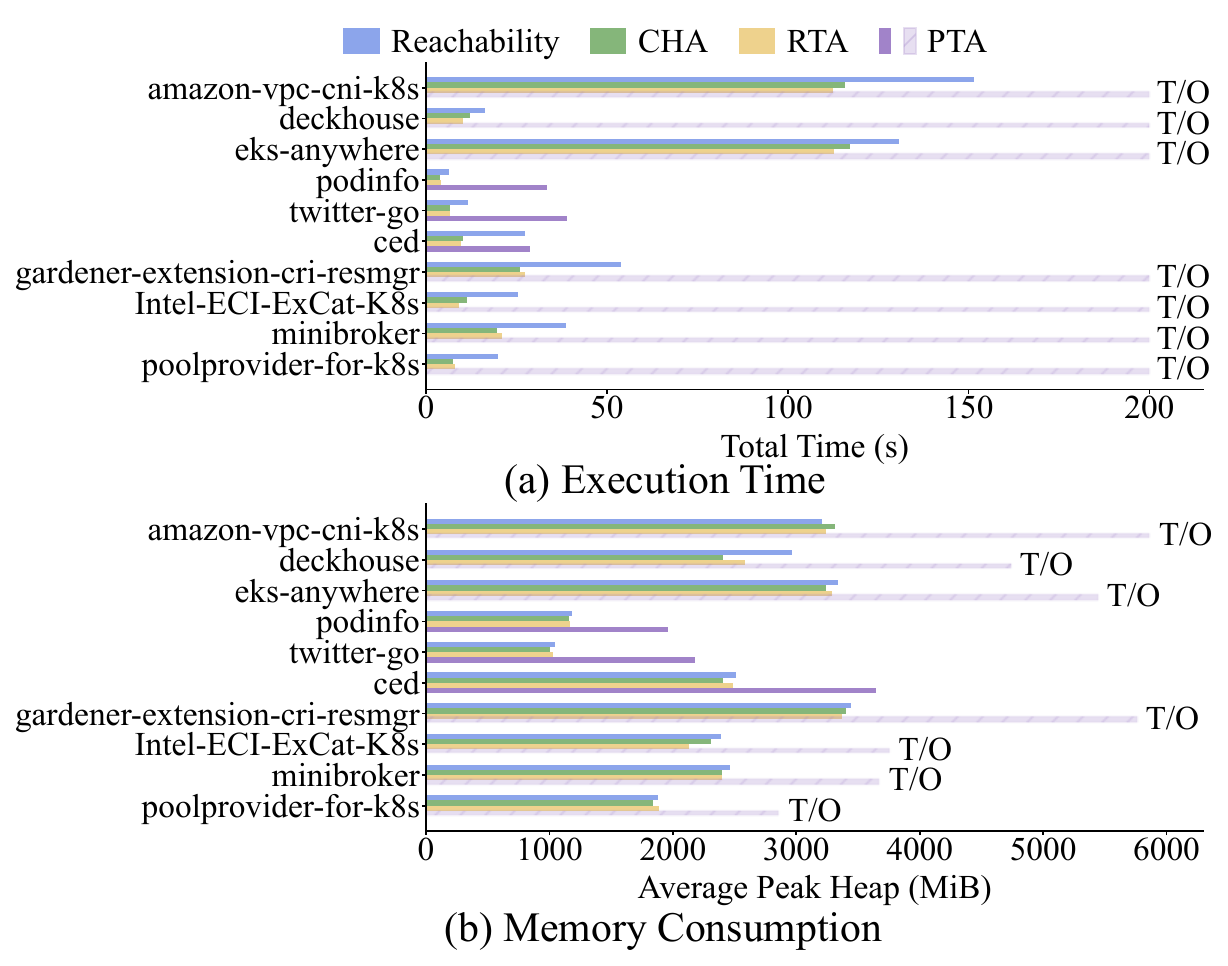}
  \caption{Execution time and memory consumption across different methods (T/O indicates a timeout after one hour).}
  \label{fig:efficiency}
\end{figure}

Figure~\ref{fig:efficiency}~(a) reports the total analysis time of four strategies, i.e., Reachability, CHA, RTA, and PTA, across 10 projects. Overall, Reachability incurs the highest runtime among the strategies that complete on all subjects, but its cost remains practical for offline analysis. In particular, the largest overhead is observed on \textit{amazon-vpc-cni-k8s} and \textit{eks-anywhere}, where Reachability takes around 150 seconds and 130 seconds, respectively. For medium-sized subjects such as \textit{gardener-extension-cri-resmgr} and \textit{minibroker}, the analysis typically finishes in about 40 to 55 seconds, while for smaller projects such as \textit{podinfo}, \textit{twitter-go}, and \textit{poolprovider-for-k8s}, it remains within roughly 5 to 20 seconds. Compared with CHA and RTA, the additional runtime of Reachability is expected, because it performs a more precise entrypoint-guided analysis rather than relying on coarser call-graph approximations. At the same time, PTA introduces the highest runtime overhead among all compared strategies and fails to complete on most projects within the one-hour timeout. Overall, these results indicate that \tool’s reachability-based design trades moderate additional time for substantially stronger capability reduction, while still completing within a few minutes on the largest subjects and within tens of seconds on most others.
% Moreover, even on the few projects where it finishes, its capability reduction results still do not surpass Reachability.

% \begin{figure}
% \begin{center}
% \includegraphics[width=\hsize]{"figures/Time"}
% \end{center}
% \caption{\label{fig:time} Execution Time across Different Methods. (T/O indicates a timeout after one hour.)}
% \end{figure}

% \begin{figure}
% \begin{center}
% \includegraphics[width=\hsize]{"figures/Memory"}
% \end{center}
% \caption{\label{fig:memory} Memory Consumption across Different Methods. (T/O indicates a timeout after one hour.)}
% \end{figure}

% \begin{figure}[htbp]
% \centering  %居中
% \subfigure[\label{fig:time} Execution Time.]{ 
% % 整体大小
% \begin{minipage}{0.45\hsize}
% % \centering    %子图居中
% % 图片大小
% % \hspace{-.5in}
% \includegraphics[scale=0.3]{"figures/Time"}
% \label{fig:6_1}
% \end{minipage}
% }
% \subfigure[\label{fig:memory} Memory Consumption.]{ 
% % 整体大小
% \begin{minipage}{0.45\hsize}
% % \centering    %子图居中
% % \hspace{-.45in}
% % \vspace{-.15in} %上下调整
% \includegraphics[scale=0.3]{"figures/Memory"}
% \label{fig:6_2}
% \end{minipage}
% }

% \caption{Execution Time and Memory Consumption across Different Methods. (T/O indicates a timeout after one hour.)}    %大图名称
% \label{evolution_sec_b}    %图片引用标记
% \end{figure}

\subsubsection{Memory overhead}
Figure~\ref{fig:efficiency}~(b) presents the average peak heap usage of four analysis strategies, i.e., Reachability, CHA, RTA, and PTA, across 10 projects. Overall, Reachability, CHA, and RTA fall into a broadly similar memory range, with most subjects requiring approximately 1 to 3.5~GiB of heap memory. The largest memory overhead is observed on \textit{amazon-vpc-cni-k8s}, \textit{eks-anywhere}, and \textit{gardener-extension-cri-resmgr}, where these three methods all consume more than 3~GiB on average. In contrast, smaller projects such as \textit{podinfo} and \textit{twitter-go} require only around 1~GiB. Compared with CHA and RTA, Reachability does not introduce a substantial memory penalty. Its peak heap usage is generally close to that of the other two methods, and in some cases even slightly lower than CHA, such as on \textit{amazon-vpc-cni-k8s}. On projects such as \textit{eks-anywhere}, \textit{deckhouse}, and \textit{gardener-extension-cri-resmgr}, the differences also remain relatively small. In contrast, PTA exhibits the highest memory overhead among all compared strategies and fails to complete on most projects. Even on the few projects where PTA finishes, its memory usage is not lower than the other methods, while its reduction effectiveness still does not surpass Reachability. Overall, these results suggest that the stronger capability reduction achieved by Reachability mainly comes at the cost of additional analysis time, rather than a significant increase in memory consumption, making \tool practical for offline analysis of real-world Kubernetes projects.

\begin{rqanswer}
\textbf{Answer to RQ3:} \tool scales practically to real-world cloud-native projects. Reachability completes within a few minutes on the largest projects and within tens of seconds on most others, while keeping memory usage broadly within 1 to 3.5~GiB.
\end{rqanswer}

% \begin{table}[t]
% \centering
% \caption{Performance Breakdown of the \tool Pipeline (in seconds)}
% \label{tab:performance}
% \resizebox{\columnwidth}{!}{%
% \begin{tabular}{@{}lrrrr@{}}
% \toprule
% \textbf{Project Name} & \textbf{Parsing \& Entry} & \textbf{IR \& Reachability} & \textbf{Data-Flow \& LLM} & \textbf{Total Time} \\ \midrule
% Nginx-ingress         & 1.2                       & 45.3                        & 12.5                      & 59.0                \\
% Prometheus-operator   & 2.5                       & 32.1                        & 8.4                       & 43.0                \\
% Argo-cd               & 3.8                       & 112.4                       & 25.6                      & 141.8               \\ \midrule
% \textbf{Average}      & \textbf{2.1}              & \textbf{52.8}                       & \textbf{15.2}                     & \textbf{70.1}               \\ \bottomrule
% \end{tabular}%
% }
% \end{table}

% \subsection{Discussion}
% \TODO{XXX}

\section{Threats to Validity}
\tool has several threats to validity. First, our current implementation focuses on Go-based Kubernetes-related projects because entrypoint recovery, SSA construction, and the subsequent static analysis are built on the Go toolchain. This limits the direct generalizability of the current system. Instead, the methodology of \tool is language-agnostic in principle. Second, highly dynamic behaviors, reflective execution, shell-heavy startup logic, and indirect argument construction may still affect analysis precision, leading to over-approximation or missed capability requirements. We currently mitigate this threat through heuristic rules and manual confirmation of uncertain cases, but a more automated solution remains needed; future work could integrate LLM-based semantic reasoning to better interpret such complex behaviors. Third, \tool employs the one-shot learning for syscall--parameter--capability rule specification, which may introduce hallucinations or inaccurate inferences. To reduce this risk, we perform human verification on LLM-assisted outputs before incorporating them into the final analysis pipeline.

%% file: tex/Section7-Conclusion.tex
\section{Conclusion}
In this paper, we first studied the prevalence of over-privileged capabilities in real-world Kubernetes projects and found that least-privilege capability configuration is still uncommon in practice. 74.67\% of projects do not explicitly configure capabilities, and only 19.33\% drop all capabilities in the datasets. To address this problem, we proposed \tool, a framework for Kubernetes capability minimization. \tool translates deployment specifications into deterministic manifests, locates actual container entrypoints, performs reachability-guided system call analysis, and leverages LLM-assisted rule specification to derive syscall--parameter--capability relations from Linux kernel code. Based on these results, \tool infers the minimal capability set required by each workload and automatically generates repaired manifests. Our evaluation on 10 representative Kubernetes-related projects shows that \tool can effectively reduce redundant capabilities in real-world workloads. On average, it achieves a capability reduction rate of 54.97\%, and its reachability-based design outperforms both RTA and CHA while maintaining practical time and memory overhead. These results show that \tool provides a practical and effective approach toward least-privilege capability enforcement in Kubernetes.

%% file: tex/Section-Acknowledgement.tex
\section{Acknowledgement}
%This work was supported by the National Natural Science Foundation of China (under Grants 62572258, 62441227, U22B2027), the Joint Fund of the National Natural Science Foundation of China (under Grant U25B2028), the Key Program of the National Natural Science Foundation of China (under Grant 62432012), the National Natural Science Foundation of Tianjin, China (under Grant 25JCQNJC01380), and the Fundamental Research Funds for the Central Universities (under Grants 079-63263257 and 079-63261161).

This work was supported by the National Natural Science Foundation of China (under Grants 62572258 and 62502237), the National Cryptologic Science Fund of China (under Grant 2025NCSF01010), the Key Program of the National Natural Science Foundation of China (under Grants U25B2028 and 62432012), and the Fundamental Research Funds for the Central Universities (under Grant 079-63263257).

%%% 贾老师
% National Natural Science Foundation of China (under Grants 62572258), the Fundamental Research Funds for the Central Universities (under Grants 079-63263257）

%%% 刘老师
% This work was supported by the National Cryptologic Science Fund of China under Grant 2025NCSF01010, and the Key Program of the National Natural Science Foundation of China under Grants U25B2028 and 62432012

%%% 维杰老师
% Weijie Liu is supported by the National Natural Science Foundation of China under Grant No.62502237.

%% file: tex/Section-Appendix.tex
\appendices

\section{Reliability Analysis of Conditional Syscall--Capability Rules}
\label{app:A}
\reviseLiu{In addition to the unconditional capability mapping, we further construct conditional syscall--capability rules to improve the precision of capability inference. The unconditional mapping records coarse-grained relations between capabilities and syscalls, covering 40 capabilities, 124 syscalls, and 164 capability--syscall mappings. However, such mapping only indicates that a syscall may require a capability, without specifying the concrete argument values or branch predicates under which the capability check is triggered. Therefore, \tool uses LLM-assisted rule specification to extract conditional rules from Linux kernel code. We manually inspected 97 conditional rules extracted from Linux 5.4, covering 41 syscalls and 21 Linux capabilities. For each rule, we checked whether the inferred capability and the direction of the guarding condition are consistent with the kernel implementation. We also checked whether the condition can be explained by directly resolvable constants or literals defined in Linux 5.4. Finally, we classified whether the original condition can be preserved or whether it must be conservatively approximated.}

\reviseLiu{Table~\ref{tab:conditional-rule} summarizes the audit results. Among the 97 extracted conditional rules, 77 rules are valid, meaning that their capability mappings and condition directions are consistent with the kernel implementation. In addition, 73 rules contain conditions that can be solved through constants or literals, and 64 rules can be resolved without over-approximation. These results show that a large portion of conditional syscall--capability rules can be handled through explicit argument-level reasoning, while the remaining cases require conservative treatment when their conditions cannot be statically resolved. As an illustrative case, consider \texttt{fcntl(int fd, int cmd, int arg)}. Different values of the \texttt{cmd} argument lead to different capability requirements. For example, \texttt{F\_SETFL} with \texttt{O\_NOATIME} may require \texttt{CAP\_FOWNER}; \texttt{F\_SETLEASE} requires \texttt{CAP\_LEASE} when setting a file lease; and \texttt{F\_SETPIPE\_SZ} may require \texttt{CAP\_SYS\_RESOURCE} when enlarging a pipe beyond the allowed limit. These cases are suitable for constant/literal-based resolution because the command values, such as \texttt{F\_SETFL}, \texttt{F\_SETLEASE}, and \texttt{F\_SETPIPE\_SZ}, are defined as kernel constants. Therefore, when a Go or C program invokes \texttt{fcntl} with these concrete command values, \tool can match the corresponding conditional rule and infer the required capability more precisely than treating all \texttt{fcntl} invocations uniformly.}

\begin{table}[!t]
\centering
\caption{Summary of the conditional syscall--capability rule audit.}
\label{tab:conditional-rule}
\renewcommand{\arraystretch}{1.15}
\begin{tabular}{lrr}
\hline
\bfseries Metric & \bfseries Count & \bfseries Percentage \\
\hline
\multicolumn{3}{l}{\itshape Extracted} \\
\#Conditional rules & 97 & -- \\
\#Syscalls & 41 & -- \\
\#Capabilities & 21 & -- \\
\hline
\multicolumn{3}{l}{\itshape Verified} \\
\#Valid rules & 77 & 79.38\%~(77/97) \\
\#Constant/Literal solved rules & 73 & 75.26\%~(73/97) \\
\#Precisely resolved rules & 64 & 65.98\%~(64/97) \\
\hline
\end{tabular}
\end{table}

\section{Detailed case study: amazon-vpc-cni-k8s}\label{app:B}

We deployed the plugin in a self-built AWS Kubernetes cluster and constructed two test Pods (\textit{nginx-pod} and \textit{nginx-pod2}) to check whether basic Pod-to-Pod communication remained functional after privilege reduction. As a control setting, we first deployed the same two Pods in a cluster using the Calico plugin, where the two Pods could successfully communicate with each other through ICMP ping. We then repeated the experiment in a cluster using Amazon VPC CNI.

We manually removed different privilege settings from the \textit{aws-node} component to examine their actual necessity. The results clearly show that not all declared privileges are equally required. First, removing all capabilities caused the component to fail its readiness probe, indicating that some privileges are indeed essential. Second, removing the \texttt{CAP\_NET\_ADMIN} capability from \textit{aws-node} also led to repeated readiness probe failures, showing that this capability is required for the correct operation of the plugin. In contrast, removing \texttt{CAP\_NET\_RAW} from \textit{aws-node} did not affect the normal execution of the component: the Pods were created successfully, the CNI plugin remained healthy, and ICMP communication between the two Pods was still preserved. This result suggests that \texttt{CAP\_NET\_RAW} is over-privileged for this workload and can be safely removed. 

We further examined another component, \textit{aws-eks-nodeagent}. After removing its declared \texttt{CAP\_NET\_ADMIN} capability, the component still functioned correctly and Pod-to-Pod connectivity remained unaffected. This provides additional evidence that some capabilities declared in the original manifest are not functionally required by the actual runtime behavior. Finally, we tested whether the coarse-grained \texttt{privileged: true} setting could be removed from \textit{aws-node}. In this case, the component entered ``\textit{CrashLoopBackOff}'', indicating that privileged execution is still necessary for certain operations in the current implementation.

% \section{Prompt example}

% \begin{myframebox}{Task Instructions}
% \begin{itemize}[
%   label=$\bullet$,
%   leftmargin=*,
%   itemsep=0.3em
% ]
%     \item \TODO{XXX}

% \end{itemize}
% \end{myframebox}